\documentclass[aps,prl,twocolumn,superscriptaddress,amsmath,nofootinbib,10pt,floatfix]{revtex4-1}
 
\usepackage{graphicx}
\usepackage[normalem]{ulem}
\usepackage{xcolor}
\usepackage[utf8]{inputenc}
\usepackage{comment}
\DeclareMathSizes{8}{7}{7}{7}

\usepackage{textgreek}
\usepackage[colorlinks, linkcolor=blue, citecolor=blue]{hyperref}
\usepackage{gensymb}
\usepackage[compat=1.1.0]{tikz-feynman}
\usepackage{tikz}
\tikzset{
pattern size/.store in=\mcSize, 
pattern size = 5pt,
pattern thickness/.store in=\mcThickness, 
pattern thickness = 0.3pt,
pattern radius/.store in=\mcRadius, 
pattern radius = 1pt}

\usepackage{amssymb}
\newcommand{\JH}{J_\tn{H}}
\newcommand{\JA}{J_\tn{A}}
\newcommand{\be}{\begin{equation}}
\newcommand{\ee}{\end{equation}}
\newcommand{\beq}{\begin{eqnarray}}
\newcommand{\eeq}{\end{eqnarray}}
\newcommand{\ba}{\[\begin{aligned}}
\newcommand{\ea}{\end{aligned}\]}
\newcommand{\bal}{\begin{aligned}}
\newcommand{\eal}{\end{aligned}}

\newcommand{\la}{\langle}
\newcommand{\ra}{\rangle}

\renewcommand{\vec}[1]{{\bf #1}}
\renewcommand{\epsilon}{\varepsilon}

\newcommand{\avg}[1]{\left\langle #1 \right\rangle}

\renewcommand{\vec}[1]{\boldsymbol{#1}}

\def \up{\uparrow}
\def \down{\downarrow}

\def \w{{\omega}}

\def \k{{\vec{k}}}
\def \q{{\vec{q}}}

\def \r{{\bf {r}}}

\def \ra{{\rangle}}
\def \la{{\langle}}
\def \tn{\textnormal}

\def \ba{\begin{align*}}
\def \ea{\end{align*}}

\newcounter{indice}

\begin{document}
\title{Spin-Polaron Mediated Superconductivity in Doped Chern Antiferromagnets}
\author{Xuepeng Wang} 
\affiliation{Department of Physics, Cornell University, Ithaca, New York 14853, USA.}
\author{J. F. Mendez-Valderrama} 
\affiliation{Department of Physics, Cornell University, Ithaca, New York 14853, USA.}
\author{Johannes S. Hofmann} \email{jhofmann@pks.mpg.de}
\affiliation{Max-Planck-Institut f\"ur Physik Komplexer Systeme, N\"othnitzer Strasse 38, 01187 Dresden, Germany}
\author{Debanjan Chowdhury}\email{debanjanchowdhury@cornell.edu}
\affiliation{Department of Physics, Cornell University, Ithaca, New York 14853, USA.}
\date{\today}

\begin{abstract}
The study of interacting topological bands with a tunable bandwidth offers a unique platform to study the interplay of intertwined orders and emergent non-electronic excitations.  
Here we design a time-reversal symmetric and sign-problem-free electronic model with tunable Chern bands carrying valley-contrasting Chern number, interacting via competing (anti-)ferromagnetic interactions. 
Using numerically exact quantum Monte-Carlo computations, we analyze the many-body phase-diagram as a function of temperature and band filling fractions over a wide range of electronic bandwidth, interaction anisotropy, and an Ising spin-orbit coupling. At a commensurate filling of the Chern bands, the ground state hosts intra-valley ferromagnetic coherence and inter-valley antiferomagnetism, thus realizing an insulating Chern antiferromagnet (CAF). Upon doping, the ground-state  develops superconductivity, but where the low-energy charged quasiparticles are composite objects --- electrons dressed by multiple spin-flip excitations. These spin-polaron (or skyrmion) excitations persist in the presence of a weak spin-orbit coupling. In a companion article \cite{wang}, we address the emergent symmetries and low-energy field-theoretic aspects of the problem and reveal the proximity to a deconfined quantum critical point. We end by providing a general outlook towards building microscopic connections with models of interacting moir\'e materials, including twisted bilayer graphene, where many of the ingredients considered here are naturally present.   
\end{abstract}

\maketitle

{\it Introduction.-} Two remarkable examples of correlated electronic systems that have captivated researchers at the frontiers of condensed matter physics over the past few decades are quantum Hall systems \cite{sarma2008perspectives} and lightly-doped Mott insulators with emergent local moments \cite{LNW}. The study of quantum Hall phases in interacting Landau levels has revealed extraordinary topological properties \cite{halperin2020fractional}, including exotic collective excitations like skyrmions \cite{girvin2000} --- nontrivial spin textures that carry electron charge. Meanwhile, lightly doped Mott insulators have demonstrated the emergence of high-temperature superconductivity from repulsive interactions, alongside various intertwined quantum orders \cite{intertwined}. The recent discovery of a variety of moir\'e materials \cite{Andrei2021,Mak2022}, with their tunable Chern bands, offers an exciting new experimental platform that elegantly bridges these two paradigmatic systems \cite{YahuiTS,BABHF,LedwithPRX}, opening fresh avenues for exploration at their intersection.

Interactions projected onto nearly flat, isolated topological bands can drive a remarkable array of correlated phenomena. A particularly elegant --- and perhaps counterintuitive --- example is the emergence of superconductivity in isolated flat bands \cite{Bernevigreview}, made possible by the complex spatial distribution of Bloch wavefunctions and interaction-driven delocalization of Cooper pairs. This superconducting mechanism operates in an intermediate to strong-coupling regime that fundamentally transcends the conventional Bardeen-Cooper-Schrieffer framework, though a variety of recent non-perturbative theoretical investigations of flat-band superconductivity have found that the Cooper pairs can still originate from underlying electrons \cite{single_chern,peri2021fragile,Zhang2021,chiral,Bernevig21}. On the other hand, previous research has proposed that composite quasiparticles such as skyrmions (or spin-polarons, corresponding to electrons dressed by spin-flip excitations) may facilitate superconductivity when doped away from various correlated insulators in topological settings \cite{ashvin_skrymion,SS20,chatterjee22}, or near deconfined quantum critical points \cite{so5_3,assaad19,assaad21,assaad23}. Nevertheless, our microscopic understanding of how superconductivity emerges at a finite temperature from a parent metallic state beyond solvable Landau-level-like systems remains limited, particularly in contexts where controlled theoretical approaches are unavailable.

In this letter, we critically examine how topology enables spin-polarons or skyrmions (terms we use interchangeably and characterize precisely below) to emerge as the lowest energy  excitations in microscopic systems distinctly removed from Landau-level-like regimes. Furthermore, we will also investigate how the superconducting transition temperature is affected due to the underlying competing ordering tendencies when doped away from Mott insulating states at commensurate band fillings. Our analysis centers on a time-reversal symmetric model of Chern bands incorporating both spin and ``valley" degrees of freedom, governed by repulsive interactions and characterized by the following microscopic symmetries: $U(1)_{\rm{charge}}\times U(1)_{\rm{valley}}\times SU(2)_{\rm{spin}}$. Our investigation begins with a lattice model of interacting Chern bands featuring tunable bandwidth, which we analyze for a range of band-fillings and temperature. Instead of incorporating the full Coulomb interactions, we focus on a specialized limit examining the interplay between an intra-valley ferromagnetic interaction, $J_{\tn{H}}$, arising from an underlying Hund's-type coupling and inter-valley antiferromagnetic interactions, $J_{\tn{A}}$, generated through inter-valley superexchange. This carefully constructed model circumvents the notorious fermion sign-problem, enabling numerically exact solutions via determinant quantum Monte Carlo (DQMC) \cite{becca2017quantum,Blankenbecler81,alf} across an extensive parameter space, that includes the band-filling, the bandwidth and the ratio $|J_{\tn{H}}|/J_{\tn{A}}$.

\begin{figure}[htb]
\includegraphics[width=86mm,scale=1]{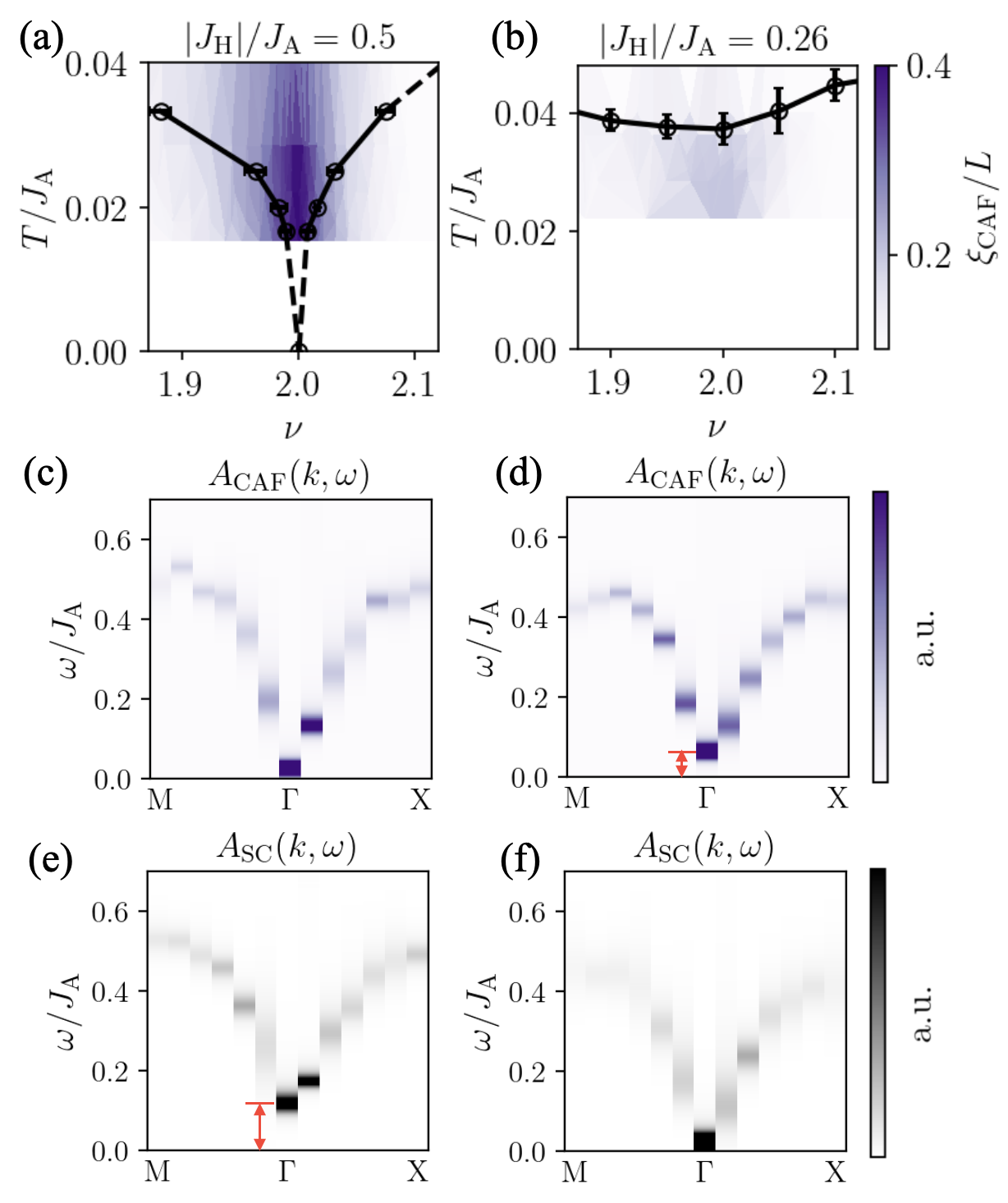}
\caption{\label{res_summary} Many-body phase diagram for $\mathcal{F}=0.01$ at $\nu=2$ with: (a) $|\JH|/\JA=0.5$, and (b) $|\JH|/\JA=0.26$. Black data points represent $T_c^{\rm{SC}}$ obtained using the BKT criterion \cite{nelson_kosterlitz} and scaling analysis \cite{si}; solid lines serve as a guide to the eye. The purple shading represents the normalized CAF correlation length, $\xi_{\rm{CAF}}/L$. Two-particle spectral function, $A_\lambda(\k,\w)$, at $\nu=2$ and $T=0$ obtained from stochastic analytical continuation for $\lambda\equiv\rm{CAF}$ (Eq.~\eqref{eq::caf}) for (c) $|\JH|/\JA=0.5$ shows Goldstone mode(s), and (d) for $|\JH|/\JA=0.26$ shows gapped excitation.  Similarly, for $\lambda\equiv\rm{SC}$, the spectral function shows (e) a gapped spectrum for $|\JH|/\JA=0.5$, and (f) a Goldstone mode for $|\JH|/\JA=0.26$. Red arrows in (d) and (e) denote the  gap in the two-particle spectra.}
\end{figure}

We have observed that the system exhibits a strong tendency to spontaneously break the $SU(2)_{\rm{spin}}$ symmetry at half-filling of the isolated electronic bands, reminiscent of quantum Hall ferromagnetism. Notably, this occurs despite fundamental differences from the spinful Landau-level problem --- our bands are not flat, the berry curvature distribution in momentum-space is non-uniform, the quantum geometry deviates from the ideal trace condition, and the interaction strength is comparable in magnitude to the gap to remote bands. The resulting ground state manifests as an insulating intra-valley Chern ferromagnet, with time-reversed valleys ordered in a relative antiferromagnetic configuration. Upon doping, electronic charge enters each valley as spin-polarons, while intervalley antiferromagnetic exchange binds these charge $e$ spin-polarons into Cooper pairs, yielding superconductivity \cite{ashvin_skrymion}. Our investigation reveals the effect of varying electronic bandwidth and the relative ratio $|J_{\tn{H}}|/J_{\tn{A}}$ on the stability of the insulating Chern antiferromagnet, the energetics of charged excitations, and superconductivity. We also present results obtained using DQMC on the collective mode spectra and the impact of spin-orbit coupling on the spin-polaron phenomenology and superconductivity. In a companion article \cite{wang} we address the low-energy field theoretic structure --- including an emergent $SO(5)$ symmetry and a proximity to deconfined quantum criticality ---alongside Hartree-Fock treatment for the same microscopic model, revealing complementary insights on many of the fundamental questions raised here.

{\it Model.-} We work with a time-reversal symmetric model of interacting topological bands with Chern number, $C=\pm1$ in two-dimensions. Microscopically, the degrees of freedom consist of spin ($\sigma=\uparrow,\downarrow$), a fictitious ``valley" ($\tau=\pm$) and sublattice ($\eta = \tn{A},\tn{B}$), such that the non-interacting part of the Hamiltonian per valley, $H^{(\tau)}_{\tn{kin}}$, in momentum space is of the form \cite{hkin,single_chern,svh},
\beq\label{h_kin0}
H^{(\tau)}_{\tn{kin}} = \sum_\k\psi^{\dagger}_{\k,\tau}\bigg[B^{0}_{\k,\tau}\eta_0 + \vec{B_{\k,\tau}\cdot\vec{\eta}}\bigg]\sigma_0\psi_{\k,\tau},
\eeq
where $\psi^{\dagger}_{\k,\tau} = (c^{\dagger}_{\k,A,\tau,\uparrow}, c^{\dagger}_{\k,B,\tau,\uparrow}, c^{\dagger}_{\k,A,\tau,\downarrow}, c^{\dagger}_{\k,B,\tau,\downarrow})$ and $c^{\dagger}_{\k,\eta,\tau,\sigma}$ denotes the electron creation operator on sublattice $\eta$ with spin $\sigma$ and valley $\tau$. The matrices $B^{0}_{\k,\tau}$ and $\vec{B}_{\k,\tau}$ are fixed by the tight-binding parameters on an underlying lattice. For simplicity, we assume a square lattice model with a first ($t$) and staggered second neighbor hopping ($t_2=t/\sqrt{2}$) with a $\pi-$flux per square plaquette \cite{si}. In order to tune the electronic bandwidth, we can include further (e.g. a fifth, $t_5$) neighbor hopping, which helps tune the flatness ratio ${\cal{F}}=W/E_{\tn{gap}}$ ($W\equiv$ bandwidth, $E_{\tn{gap}}\equiv$ remote bandgap) to be small. We construct two copies of $H^{(\tau)}_{\tn{kin}}$ in a time-reversal-invariant fashion \cite{si}, under the operation $\mathcal{T}=i\tau_y\mathcal{K}$ where $\mathcal{K}$ denotes complex conjugation. This yields a model with a set of degenerate topological bands carrying spin and valley with $C=\tau$. 

In our previous discussion, we presented two competing local magnetic interactions: a ferromagnetic intra-valley Hund's interaction with $J_{\tn{H}}<0$ and an antiferromagnetic inter-valley exchange with $J_{\tn{A}}>0$. We will examine how these interactions once effectively projected to the low-energy Chern bands influence the competition between spontaneously symmetry-broken phases at commensurate fillings and superconductivity in our proposed model. While these interaction forms draw partial inspiration from quantum Hall ferromagnetism in spinful Landau levels \cite{gmp,sondhi,qhbilayer,girvin1999quantum}, our framework operates well beyond any readily ``solvable" quantum Hall regime. The interactions take the form, $H_{\tn{interaction}} = H_{\tn{intra-valley}} + H_{\tn{inter-valley}}$, where
\begin{subequations}\label{eq::ham_int}
\beq
H_{\tn{intra-valley}} &=& 
 - |J_{\tn{H}}|\sum_{\r,\tau=\pm} \vec{S}^{\tau}_{\r}\cdot\vec{S}^{\tau}_{\vec{\r}}, \label{intra_valley_int}\\
H_{\tn{inter-valley}} &=& J_{\tn{A}}\sum_{\vec{r}}\vec{S}^{+}_{\r}\cdot\vec{S}^{-}_{\r},~~\tn{where}\label{inter_valley_int}\\
\vec{S}^{\tau}_{\r} &=& \sum_{\alpha,\beta = \uparrow,\downarrow}c^{\dagger}_{\r,\tau,\alpha}\vec{\sigma}_{\alpha\beta}c_{\r,\tau,\beta}.
\eeq
\end{subequations}
Note that we combine the sublattice index ($\eta$) into the two-dimensional spatial coordinate $\r$. 
\begin{figure*}[htb]
\includegraphics[width=175mm,scale=1]{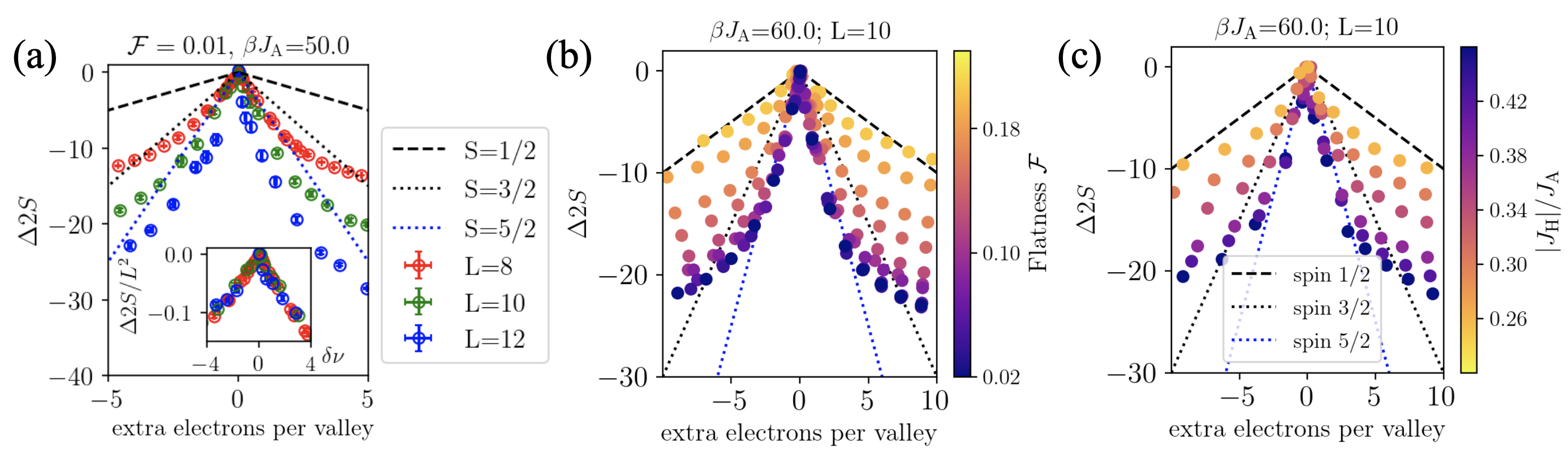}
\caption{\label{Fig::skyr_fig2} Evolution of spin quantum number, $\Delta S \equiv S^\tau_{\tn{tot}}(\nu) - S^\tau_{\tn{tot}}(\nu=2)$, within a single valley at a fixed: (a) $|\JH|/\JA=0.5$, $\mathcal{F}=0.01$ and $\beta J_{\rm{A}}=50$ as a function of increasing linear system size, $L$, (b) $|\JH|/\JA=0.5$ and $\beta J_{\rm{A}}=60$ with increasing $\mathcal{F}$, (c)  $\mathcal{F}=0.01$ and $\beta J_{\rm{A}}=60$ with decreasing $|\JH|/\JA$. Inset of panel (a) shows $(\Delta S/L^2)$, indicating the macroscopic nature of the charged skyrmion excitation. 
}
\end{figure*}

{\it Quantum Monte Carlo Simulations.-} Our proposed model features an anti-unitary time-reversal symmetry, $\mathcal{T}'=i\tau_x\sigma_y\mathcal{K}$, which allows for sign-problem-free DQMC at {\it any} filling fraction $\nu$, provided that $J_{\rm{A}}\geq 2|J_{\rm{H}}|$. We evaluate the partition function for the model defined by $H=(\sum_{\tau}H^{(\tau)}_{\tn{kin}}+H_{\tn{interaction}})$ using a Trotter decomposition with $\Delta \tau = \beta/N_{\tn{Trotter}}$; the interaction terms are factorized through a discrete Hubbard-Stratonovich transformation, with auxiliary fields sampled stochastically via single spin-flip updates. We will analyze the low temperature phase diagram and the low-energy excitation spectrum for a range of bare band dispersion, ${\cal{F}}=0.01-0.2$, and for interaction strengths $J_{\rm{A}}/E_{\rm{gap}}=0.625$, where $E_{\rm{gap}}=4t$. We determine the band-filling by adjusting the chemical potential $\mu(T)$ to satisfy $\sum_{\r}\langle n_{\r}\rangle/L^2 = \nu$ , where $\nu$ represents the total filling and $n_{\r}=\sum_{\tau,\sigma}c^{\dagger}_{\r,\tau,\sigma}c_{\r,\tau,\sigma}$ represents the local electron density ($L^2\equiv$ system size). In the parameter regime where the model is sign-problem free, the inter-valley antiferromagnetic exchange is large relative to the intra-valley ferromagnetic exchange; we focus primarily on the limit $J_{\rm{A}}=2|J_{\rm{H}}|$ in the remainder of this study where the Hunds' interaction relative to the antiferromagnetic exchange is maximized. However, we have also observed interesting quantum phase transitions as a function of decreasing $|J_{\rm{H}}|/J_{\rm{A}}$ at a fixed $\nu=2$, reminiscent of deconfined quantum criticality with emergent $SO(5)$ symmetry \cite{wang}.

{\it Phase-diagram.-} 
Let us start by discussing the many-body phase diagram for $\mathcal{F}=0.01$. The candidate order-parameters of interest are a spin-singlet valley-triplet ($\tau_z=0$) superconductivity and a Chern-antiferromagnet (CAF) defined as,
\begin{subequations}
\beq
&O_{\tn{SC}}\equiv c^{\dagger}_{\r+\up} c^{\dagger}_{\r-\down} - c^{\dagger}_{\r+\down} c^{\dagger}_{\r-\up},\label{eq::sc_spinsing}\\
&O^{\alpha}_{\tn{CAF}}\equiv \psi^{\dagger}_{\vec{r}}\sigma^{\alpha}\tau^{z}\psi_{\vec{r}}, ~~(\alpha = 1,2,3).\label{eq::caf}
\eeq
\end{subequations}
We use the Berezinskii–Kosterlitz–Thouless (BKT) criterion \cite{bkt1, bkt2} and scaling analysis to determine the superconducting transition temperature $T_c^{\tn{SC}}$ \cite{si}. The data for $T_c^{\tn{SC}}$ is shown in Fig.~\ref{res_summary}(a)-(b) for two different values of $|\JH|/\JA$ in the vicinity of $\nu=2$; the superconducting pairing symmetry is given by Eq.~\ref{eq::sc_spinsing}.

We find that for $|\JH|/\JA=0.5$, which lies at the boundary of parameter-space where the sign-problem is absent, the superconducting $T_c^{\rm{SC}}\rightarrow0$ as $\nu\rightarrow2$; see Fig.~\ref{res_summary}(a). The ground state at $\nu=2$ is an insulating CAF \cite{wang} with an order-parameter described by Eq.~\ref{eq::caf}. At a finite temperature and for $|\JH|/\JA=0.5$, the CAF correlation length, $\xi_{\tn{CAF}}$, shows a strong enhancement upon approaching $\nu=2$ as a function of decreasing temperature, as shown in Fig.~\ref{res_summary}(a). However, the CAF at $\nu=2$ is restricted to only a certain range of accessible $|\JH|/\JA$. For instance, when $|\JH|/\JA=0.26$, the superconducting $T_c^{\rm{SC}}$ is finite in the entire neighborhood of $\nu=2$; see Fig.~\ref{res_summary}(b).  Interestingly, there is still a small enhancement in $\xi_{\tn{CAF}}$ near $\nu=2$ for $|\JH|/\JA=0.26$, which has a rather weak effect on $T_c^{\rm{SC}}$, as seen in Fig.~\ref{res_summary}(b).

By performing a $T=0$ projective quantum Monte-Carlo computation, we have established that the $\nu=2$ ground state for $|\JH|/\JA=0.5$ is a CAF \cite{wang}. This is corroborated by analyzing the spectral properties (after analytic continuation to real frequencies) of two-point correlation functions involving the order-parameters introduced above. Specifically, by analyzing the frequency and momentum-resolved spectral function, $A_\lambda(\k,\omega)=\tn{Im}\la O^\dagger_\lambda(\k,\omega) O_\lambda(\k,\omega)\ra |_{\omega\rightarrow\omega+i0^{+}}~~(\lambda\equiv\rm{SC,CAF})$,  in the limit of $\k\rightarrow0$, we can identify the gapless Goldstone modes due to the associated spontaneously broken continuous symmetries. For $|\JH|/\JA=0.5$ and at $\nu=2$, we observe gapless mode(s) near the $\Gamma-$point in the CAF spectral function (Fig.~\ref{res_summary}c), while for $|\JH|/\JA=0.26$ the same mode(s) are gapped (Fig.~\ref{res_summary}d). Relatedly, for $|\JH|/\JA=0.5$ at $\nu=2$, the SC spectral function (Fig.~\ref{res_summary}e) is gapped near the $\Gamma-$point, while for $|\JH|/\JA=0.26$ the same mode is gapless (Fig.~\ref{res_summary}f). Thus, we have demonstrated that at $\nu=2$, by simply tuning the ratio $|\JH|/\JA$, we can transit from an insulating Chern antiferromagnet to a superconductor \cite{wang}.

{\it Charged excitations.-} We now examine the evolution of the ground-state when charges are introduced to the incompressible CAF at $\nu=2$. The observable of interest will be the variation in total spin quantum number, $\Delta S$, as a function of the excess electrons {\it per valley}. This is reminiscent of the Knight-shift measurements performed in the original quantum Hall setting to reveal the existence of skyrmions \cite{barrett}. Clearly, if the doped charges enter each valley as a renormalized electronic quasiparticle, the slope should reflect the fundamental 
$S=1/2$ quantum number. Conversely, if the lowest energy excitations per valley consist of composite entities involving spin-flips, this will manifest in the slope with $S>1/2$. Recall that the static spin structure factor {\it per valley} ($\tau$), $\chi^{\tau}(\q)$, can be related to the total (i.e. macroscopic) spin quantum number $S_{\tn{tot}}^{\tau}$ per valley via
\begin{subequations}
\beq
\chi^{\tau}(\q) &=& \frac{1}{L^2}\sum_{\r,\r'}e^{-i(\r-\r')\cdot\q}\avg{\vec{S}^{\tau}(\r')\cdot\vec{S}^{\tau}(\r)},\\
\chi^{\tau}(\q=0) &=& S_{\tn{tot}}^{\tau}(S_{\tn{tot}}^{\tau}+1)/L^2, 
\eeq
\end{subequations} 
where we have dropped some unimportant constant terms. We then compute, $\Delta S \equiv S_{\tn{tot}}^{\tau}(\nu) - S_{\tn{tot}}^{\tau}(\nu=2)$, defined as the change in the spin quantum number per valley with doping relative to the maximally polarized CAF state at $\nu=2$ from $\chi^\tau(\q=0)$.

In Fig.~\ref{Fig::skyr_fig2}, we analyze $\Delta S$ as a function of increasing system-size, flatness ratio, ${\cal{F}}$, and $|\JH|/\JA$. At a fixed ${\cal{F}}=0.01,~|\JH|/\JA=0.5$ and a low temperature $\beta J_{\rm{A}}=50$, we find clear evidence based on $\Delta S$ that the doped charge does {\it not} enter as a $S=1/2$ electron; see e.g. the strong deviation away from the line with slope $S=1/2$ in Fig.~\ref{Fig::skyr_fig2}(a). Instead, $\Delta S$ appears to involve combinations of $S=3/2,~5/2,..$ excitations, which in turn depend on the system size, $L\times L$. These $S>1/2$ objects are associated with the electron dressed by multiple spin-flip excitations derived from the intra-valley Chern ferromagnet. If the resulting composite excitation is tightly bound, it is appropriate to describe it as a spin-polaron. However, by studying $(\Delta S/L^2)$ as a function of the total doping summed over {\it both} valleys, $\delta\nu$, we observe that the charge enters the system as an extended macroscopic object (Inset-Fig.~\ref{Fig::skyr_fig2}a), reminiscent of a skyrmion \cite{girvin2000}. We turn next to one of the technical advantages of our setup, which allows us to examine the stability of the insulating phase at $\nu=2$, and the energetics associated with the doped charges as a function of ${\cal{F}}$ and $|\JH|/\JA$.

With increasing ${\cal{F}}$ at a fixed $|\JH|/\JA$ (Fig.~\ref{Fig::skyr_fig2}b), or decreasing $|\JH|/\JA$ at a fixed ${\cal{F}}$ (Fig.~\ref{Fig::skyr_fig2}b), we continue to observe a substantial parameter space where the 
$S>1/2$ skyrmion excitations remain the lowest energy charged excitations, highlighting their stability. The system exhibits a gradual crossover to a regime --- when the bandwidth is relatively large or the intervalley antiferromagnetic exchange dominates significantly over the intravalley ferromagnetic exchange --- where electrons once again become the lowest-energy charged excitations. Our results in Fig.~\ref{Fig::skyr_fig2}(c) also show that with decreasing $|\JH|/\JA$, the tendency towards melting the chern antiferromagnet at $\nu=2$ decreases the skyrmion size. Remarkably, at $|\JH|/\JA = 0.26$ when the ground state at $\nu=2$ is a superconductor ($T_c^{\tn{SC}}/\JA = 0.037(3)$), the doped charge still enters the system as a skyrmion thereby demonstrating that the mechanism for superconductivity is driven by pairing of a charge$-e$ skyrmion and antiskyrmion, which is also suggestive of an underlying $(|\JH|/\JA)-$tuned deconfined quantum phase transition \cite{wang}. Interestingly, we have seen preliminary tendency towards a non-exponential saturation of the temperature dependence of $\rho_s(T)$ in the lightly doped skyrmion-mediated superconductor, likely due to the presence of the gapless excitations in the proximate CAF phase \cite{si}. 

\begin{figure}[h!]
\includegraphics[width=85mm,scale=1]{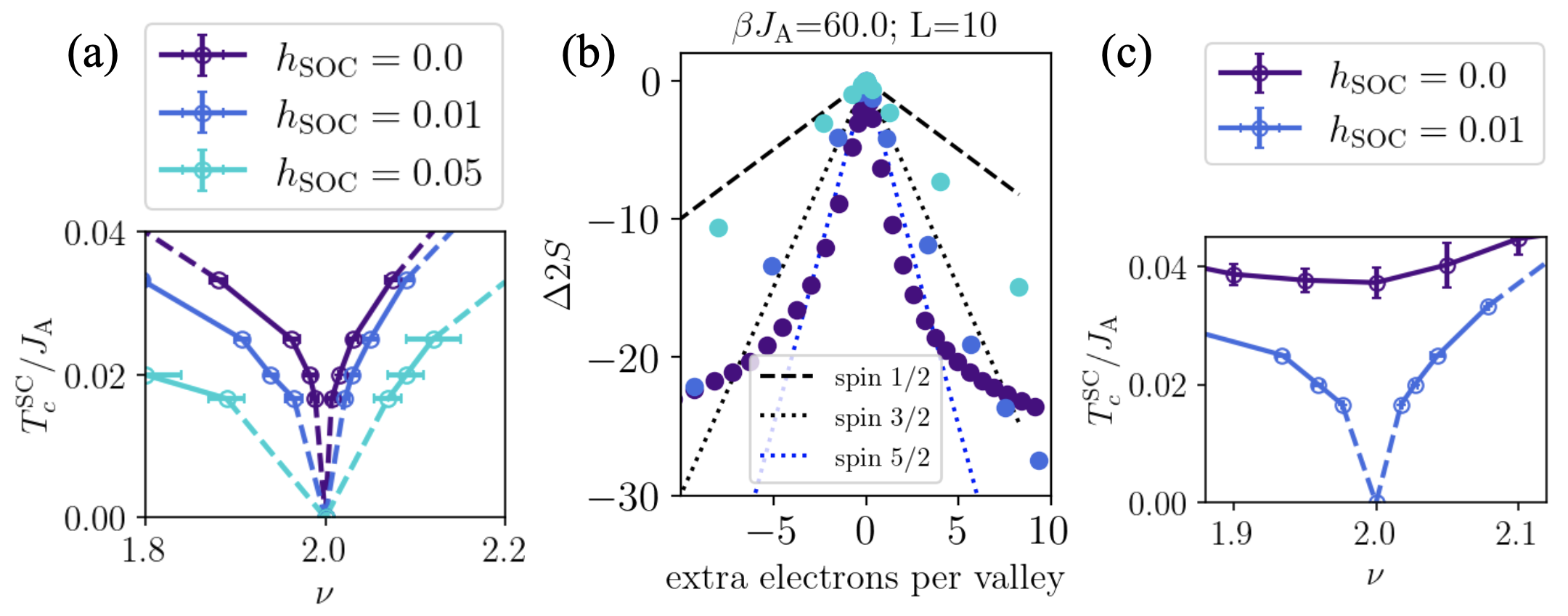}
\caption{\label{skyr_fig4} (a) Many-body phase diagram for  $\mathcal{F}=0.01$ and $|\JH|/\JA=0.5$ with increasing $h_{\rm{SOC}}$. (b) $\Delta S$ within a single valley measured relative to $\nu=2$ with increasing $h_{\rm{SOC}}$ at $\beta J_A=60.0$. (c) Evolution of $T_c^{\rm{SC}}$ with $\nu$ for $\mathcal{F}=0.01$ and $|\JH|/\JA=0.26$ for $h_{\rm{SOC}}\neq0$. }
\end{figure}

{\it Effect of spin-orbit coupling.-} Our discussion has so far focused on a model with explicit $SU(2)_{\rm{spin}}$ symmetry that is broken spontaneously in the CAF phase at $\nu=2$, leading to the emergence of skyrmion excitations at $\nu=2+\delta\nu$. This naturally raises the question of whether the skyrmions survive if the microscopic $SU(2)_{\rm{spin}}$ symmetry is explicitly broken, and how  it affects $T_c^{\rm{SC}}$. One way to address this question is to include a spin-orbit coupling term at the single-particle level in the Hamiltonian, $H\rightarrow H + H_{\rm{SOC}} $,  
\beq
H_{\rm{SOC}} =  h_{\rm{SOC}} \sum_{\r\tau;\alpha,\beta = \uparrow,\downarrow}\tau c^{\dagger}_{\r,\tau,\alpha}\sigma^{z}_{\alpha\beta}c_{\r,\tau,\beta},
\eeq
without affecting the sign-problem-free property of the full model. 

We start by investigating the effect of increasing $h_{\tn{SOC}}$ on the many-body phase-diagram in the vicinity of $\nu=2$ for $|\JH|/\JA=0.5$ and $\mathcal{F}=0.01$. We find that the insulating phase at $\nu=2$ survives, and the magnitude of $T_c^{\rm{SC}}$ obtained by doping away from $\nu=2$ is suppressed with increasing $h_{\tn{SOC}}$; see Fig.~\ref{skyr_fig4}(a). The $h_{\rm{SOC}}$ term breaks the degeneracy between the two contributions to $O_{\rm{SC}}$ in Eq.~\ref{eq::sc_spinsing} and mixes the spin-singlet and spin-triplet superconducting orders. The evolution of $\Delta S$ as a function of the doped charge per valley near $\nu=2$ is shown in Fig.~\ref{skyr_fig4}(b) as a function of increasing $h_{\rm{SOC}}$. Remarkably, the $S>1/2$ charged excitations still survive at $T\sim h_{\rm{SOC}}$, where thermal fluctuations effectively restore the weakly broken $SU(2)_{\rm{spin}}$ (due to $h_{\rm{SOC}}$). In other words, at these temperatures, the thermally excited skyrmions are the cheapest charged excitations. With increasing $h_{\rm{SOC}}$, the skyrmion size decreases culminating in an eventual crossover to ordinary electronic quasiparticles. Finally, in Fig.~\ref{skyr_fig4}(c), we present the many-body phase-diagram for $|\JH|/\JA=0.26$ and $\mathcal{F}=0.01$, where in the absence of $h_{\rm{SOC}}$ there is no CAF phase. However, with a finite $h_{\rm{SOC}}$, the SU(2)$_{\textnormal{spin}}$ degeneracy in the CAF state at $\nu=2$ is lifted by the spin-orbit coupling, leading to a spin-valley Hall band-insulator. 
Upon doping, the system develops superconductivity with $T_c^{\rm{SC}}$ suppressed compared to the $h_{\rm{SOC}}=0$ case. Developing an analytical understanding of why $T_c^{\rm{SC}}$ is systematically suppressed with increasing $h_{\rm{SOC}}$, including when the correlated CAF phase survives at $\nu=2$, remains an interesting  future direction.

{\it Outlook.-}
Starting from an exactly solvable model with interacting topological bands with a tunable bandwidth, we have demonstrated unambiguous evidence for the presence of composite charged excitations with $S>1/2$ when doped away from an interaction-induced Chern antiferromagnet at partial commensurate filling. Moreover, the same interactions that generate the insulator also drive pairing of these composite charge excitations to yield a superconductor.  We have also demonstrated that the excitations survive in the
presence of a weak spin-orbit coupling at a finite temperature. Our findings show a concrete and highly tunable setting at a finite temperature where charged skyrmions-mediated superconductivity has been proposed as a potential mechanism for moir\'e materials \cite{ashvin_skrymion, baby_skyrmion,Schindler22,chatterjee22,Yves22}. However, we note that our study  is in a regime that is far removed from any ideal Landau-level like effective description, due to the finite electronic dispersion, non-concentrated and non-ideal band geometry, and a small interaction-induced mixing with the remote bands. In a companion article \cite{wang}, we analyze the quantum phase transitions that arise when tuning the bandwidth and interaction anisotropies, and also comment on how the non-ideal band geometry affects the local charged spin-textures.

Our study is inspired by the phenomenology of moir\'e graphene and some of our findings share a broad conceptual similarity with their microscopic models \cite{tarnopolsky_origin_2019,bultinck_ground_2020,TBG4,KangVafekPRL}, although we do not capture their microscopic bandstructures or the full Coulomb repulsion. It is presently unclear if there are clear experimental signatures of skyrmion-mediated superconductivity in these materials. Interestingly, as a matter of principle, we have demonstrated that a proximity-induced inversion symmetry breaking from a nearby WSe$_2$ gate \cite{yu2022spin, stevan20} does not necessarily suppress the skyrmion excitations altogether. 
Developing a detailed theory of electrical transport above $T_c$ in the doped metallic state and of the superfluid stiffness far below $T_c$ in the superconducting state remain interesting future research directions. Finding sharp experimental signatures of these skyrmion-like excitations also remains a promising direction. 

{\it Acknowledgments.-} We thank E. Khalaf and A. Vishwanath for insightful discussions. X.W. thanks J. Dong, P. J. Ledwith and R. Sahay for useful discussions. D.C. is supported in part by a NSF CAREER grant (DMR-2237522), and a Sloan Research Fellowship. D.C. also acknowledges the hospitality of MPI-PKS, Dresden, during the final stages of preparation of this manuscript. This work used Expanse at the San Diego Supercomputer Center through allocation TG-PHY240209 from the Advanced Cyberinfrastructure Coordination Ecosystem: Services \& Support (ACCESS) program \cite{access}, which is supported by National Science Foundation grants \#2138259, \#2138286, \#2138307, \#2137603, and \#2138296.
The auxiliary field QMC simulations were carried out using the ALF package \cite{alf}.
This work was supported in part by the Deutsche Forschungsgemeinschaft DFG through the Würzburg-Dresden Cluster of Excellence on Complexity and Topology in Quantum Matter `ct.qmat' (EXC2147, project ID 390858490).

\bibliographystyle{apsrev4-1_custom}
\bibliography{draftv1.bib}

\clearpage
\renewcommand{\thefigure}{S\arabic{figure}}
\renewcommand{\figurename}{Supplemental Figure}
\setcounter{figure}{0}
\renewcommand{\theequation}{S\arabic{equation}}
\setcounter{equation}{0}
\begin{widetext}

\begin{center}
\Large{\textbf{Supplemental Material for} ``Spin-Polaron Mediated Superconductivity in Doped Chern Antiferromagnets"}\\
\end{center}

\begin{center}
\large{ X. Wang,~J.F. Mendez-Valderrama,~J.S. Hofmann,~D. Chowdhury}
\end{center}

\section{Band dispersion and Quantum Geometry}
As introduced in the main text, the matrices in Eq. \ref{h_kin0} that determine the dispersion for the free-particle bands are given by,
\begin{equation}
\begin{aligned}
B^{x}_{\k,\tau=+} + iB^{y}_{\k,\tau=+} &= -2t \bigg[e^{-i\frac{\pi}{4}-ik_y}\cos(k_y) + e^{i\frac{\pi}{4}-ik_y}\cos(k_x)\bigg] \\
B^{z}_{\k,\tau=+} &= -2t_2 \bigg[ \cos(k_x + k_y) - \cos(k_x - k_y) \bigg] \\
B^{0}_{\k,\tau=+} &= -2t_5 \bigg[ \cos(2k_x + 2k_y) + \cos(2k_x - 2k_y) \bigg]
\end{aligned},
\end{equation}
where $t,~t_2,~t_5$ are the hopping parameters. For the simulations with $\mathcal{F}=0.01$, the hopping parameters are given by $t_2 = t/\sqrt{2}$ and $t_5 = (1-\sqrt{2})t/4 $ \cite{single_chern}; whereas for the simulations with $\mathcal{F}=0.2$, the hopping parameters are given by  $t_2 = t/\sqrt{2}$ and $t_5 = 0 $, respectively. 
The non-interacting band structure, associated Berry curvature distribution and deviation from ideal band geometry in the Brillouin zone are shown in Fig.~\ref{Fig::band}.

\begin{figure*}[htb]
\includegraphics[width=160mm,scale=1]{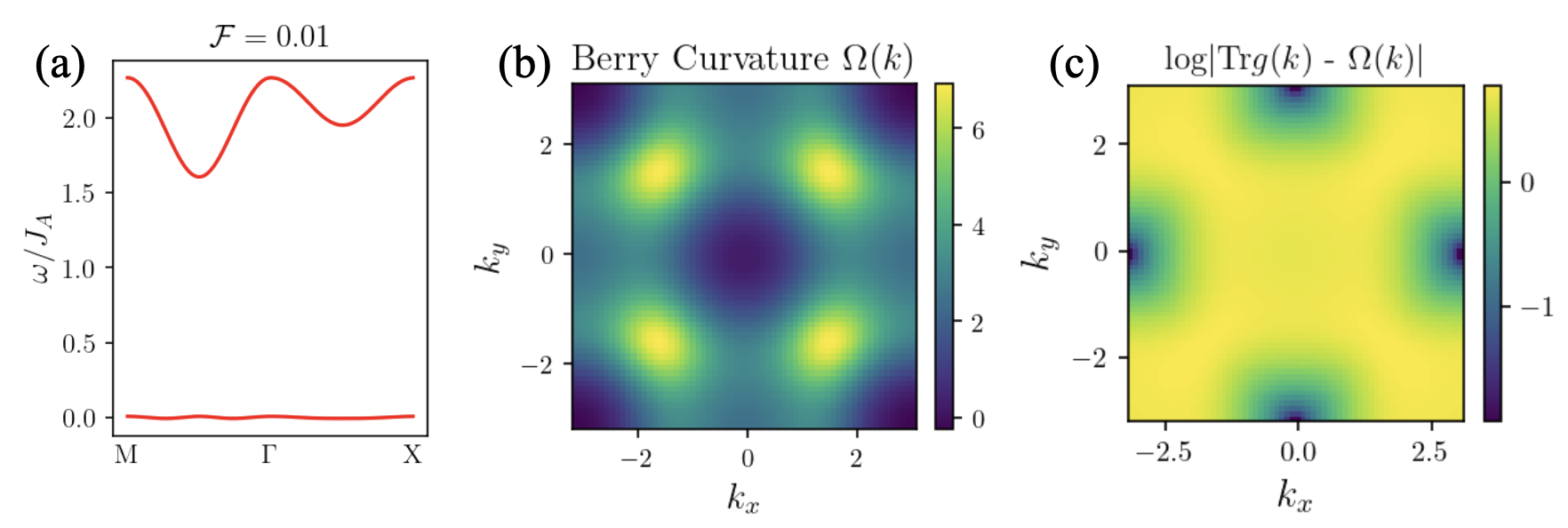}
\caption{\label{Fig::band} 
 (a) Non-interacting band structure along a high-symmetry cut in the Brillouin zone for $\mathcal{F} = 0.01$. (b) Berry curvature distribution $\Omega(\k)$, which is independent of $\mathcal{F}$ since $t_5$ couples to the identity matrix in sublattice space. (c) Deviation from ideal band geometry in a log-scale. Here $g(\k)$ denotes the quantum geometric tensor.
}
\end{figure*}

\section{Sign-Problem-Free simulation}\label{App::sign problem}
In this section, we discuss the origin of the sign-problem-free nature of the model defined in Eqn.~\ref{h_kin0} and \ref{eq::ham_int}, respectively. For the auxiliary field quantum Monte-carlo simulations, the presence of an anti-unitary symmetry, $\mathcal{T}' = i\tau_x \sigma_y\mathcal{K}$, is key \cite{congjun}. Let us re-write the interaction  Hamiltonian, $H_{\tn{interaction}}$, in Eqn.~\ref{eq::ham_int} as,
\beq
H_{\tn{interaction}} &= \sum_{\vec{r}} \bigg[J_1 (\vec{S}^{+}_{\vec{r}} + \vec{S}^{-}_{\vec{r}})^2 - J_2 (\vec{S}^{+}_{\vec{r}} - \vec{S}^{-}_{\vec{r}})^2 \bigg],
\eeq
where $J_1 = J_{\tn{A}}/4 + J_{\tn{H}}/2$ and $J_2 = J_{\tn{A}}/4 - J_{\tn{H}}/2$, respectively. Introducing a Hubbard-Stratonovich decomposition using auxiliary fields, $\vec{\phi}_{1,\r}$, $\vec{\phi}_{2,\r}$, in the spin channel, at leading order in $\Delta \tau$, the imaginary-time evolution operator is given by,
\beq
-\Delta\tau H_{\tn{interaction}} &= \sum_{\vec{r}} \bigg[\sqrt{-\Delta\tau J_1} ~ \vec{\phi}_{1,\r} \cdot (\vec{S}^{+}_{\vec{r}} + \vec{S}^{-}_{\vec{r}}) - i \sqrt{-\Delta\tau J_2} \vec{\phi}_{2,\r} \cdot(\vec{S}^{+}_{\vec{r}} - \vec{S}^{-}_{\vec{r}})\bigg].
\eeq
The spin operators in each valley transform under $\mathcal{T}'$ as $\mathcal{T}' \vec{S}^{\pm}_{\vec{r}} \mathcal{T}'^{-1} = -\vec{S}^{\mp}_{\vec{r}}$. It is straightforward to show that if $J_1>0$ and $J_2>0$, $(-\Delta\tau H_{\tn{interaction}})$ is invariant under $\mathcal{T}'$, which ensures the eigenstates of $H_{\tn{interaction}}$ can be grouped into Kramer doublets, guaranteeing the positive-definiteness of the partition function \cite{congjun}. Similar argument also hold for $H_{\tn{kin}}$, and thus the model is sign-problem-free.

\section{Supplementary Data for Determination of $T_c^{\rm{SC}}$}\label{App::supp_data}
To characterize the onset of superconductivity, we compute the correlation length,
\begin{subequations}
\beq
&
\xi_O \equiv \frac{1}{2 \sin(\pi/L)}\sqrt{\frac{S_O(\vec{q} = \vec{0})}{S_O(\vec{q} = (2\pi/L,0))} - 1},\\&
S_O(\q) = \frac{1}{L^2}\sum_{\r,\r'}e^{-i(\r-\r')\cdot\q}\avg{O^{\dagger}(\r')O(\r)},
\eeq
\end{subequations}
where $O$ is the superconducting order parameter defined in Eq.~\ref{eq::sc_spinsing}. We also compute the temperature-dependent superfluid stiffness, $\rho_s(T)$, as the transverse electromagnetic response at vanishing Matsubara frequency \cite{criterion},
\beq\label{eq::stiffness}
\rho_s = \frac{1}{4}\avg{[-K_{xx}-\Lambda_{xx}(\omega_n, q_y=0, q_x \rightarrow 0)]},
\eeq
where $\Lambda_{xx}(\omega_n, \vec{q})$ is paramagnetic current-current correlation, and $K_{xx} \equiv\avg{\partial^2 H[\vec{A}]/\partial A_{x}^2|_{\vec{A}\rightarrow0}}$ is the diamagnetic contribution, with $\vec{A}$ the probe vector potential. The superconducting transition temperature, $T_c^{\tn{SC}}$, is then determined as $T_c^{\tn{SC}} = \pi \rho_s (T\rightarrow T_c^{-})/2$ \cite{bkt1, bkt2}.

We extract the superconducting critical filling $\nu_c$ at a fixed inverse temperature $\beta$ for $|\JH|/\JA=0.5$ through scaling analysis of $\xi_{\rm{SC}}/L$ for $h_{\tn{SOC}}=0$ (Fig.~\ref{Fig::hsoc0}) and for $h_{\tn{SOC}}=0.01$ (Fig.~\ref{Fig::hsoc001}). The superconducting critical filling $\nu_c$ at fixed inverse temperature $\beta$ for $h_{\tn{SOC}}=0.05$ are extracted based on BKT criterion, as shown in Fig.~\ref{Fig::hsoc005}.

\begin{figure*}[htb]
\includegraphics[width=180mm,scale=1]{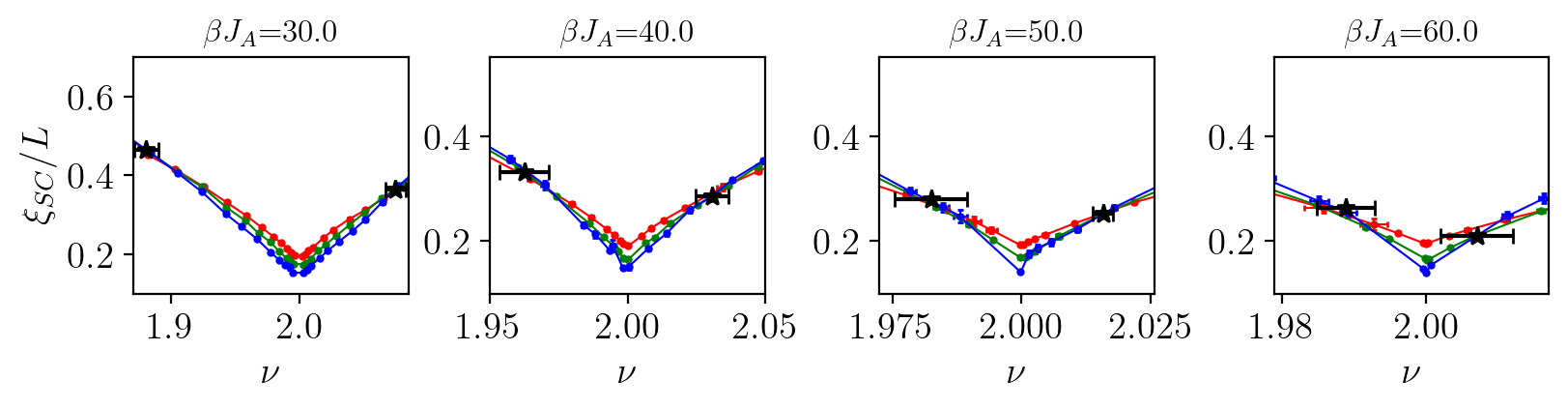}
\caption{\label{Fig::hsoc0}
Superconducting correlation length $\xi_{\rm{SC}}$ as a function of $\nu$ for $\mathcal{F}=0.01$, $|\JH|/\JA=0.5$ and $h_{\tn{SOC}}=0$. Black star with errorbar denotes the critical filling $\nu_c$ at the respective temperature. Color schemes are the same as that in Fig.\ref{Fig::skyr_fig2}a.
}
\end{figure*}

\begin{figure*}[htb]
\includegraphics[width=180mm,scale=1]{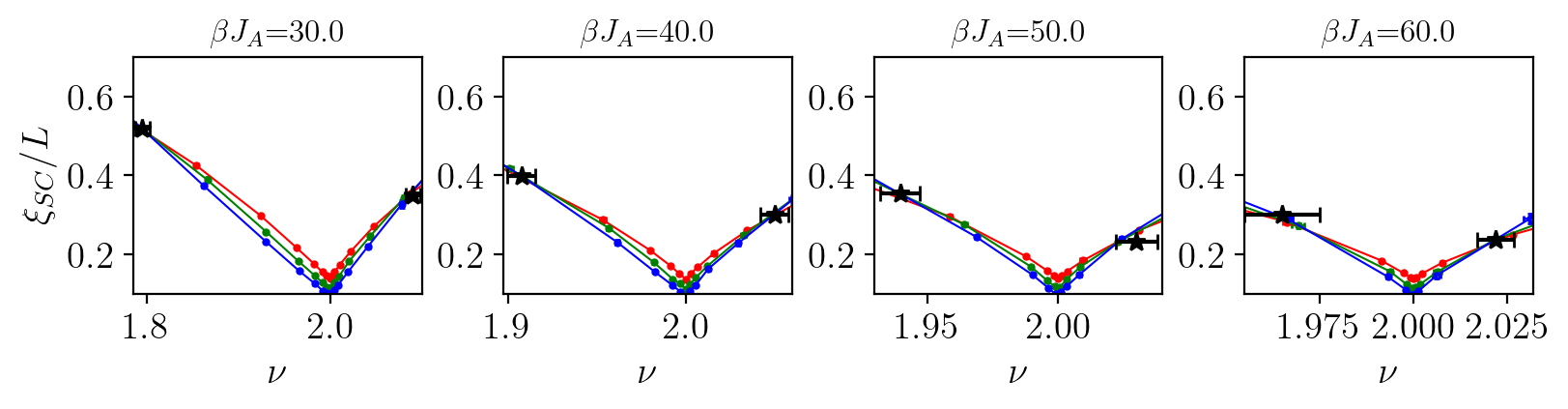}
\caption{\label{Fig::hsoc001} 
Superconducting correlation length $\xi_{\rm{SC}}$ as a function of $\nu$ for $\mathcal{F}=0.01$, $|\JH|/\JA=0.5$ and $h_{\tn{SOC}}=0.01$. Black star with errorbar denotes the critical filling $\nu_c$ at the respective temperature. Color schemes are the same as that in Fig.~\ref{Fig::skyr_fig2}a
}
\end{figure*}

\begin{figure*}[htb]
\includegraphics[width=140mm,scale=1]{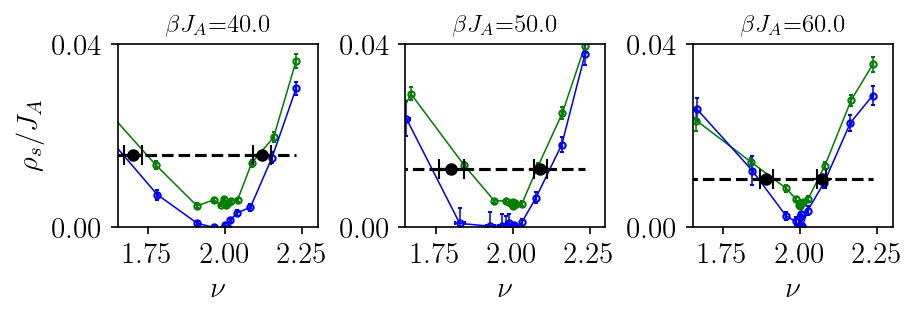}
\caption{\label{Fig::hsoc005} 
Superfluid stiffness $\rho_s$ as a function of varying filling fraction $\nu$, calculated through Eq.~\ref{eq::stiffness} for $\mathcal{F}=0.01$, $|\JH|/\JA=0.5$ and $h_{\tn{SOC}}=0.05$. Color schemes are the same as that in Fig.~\ref{Fig::skyr_fig2}a. Black circle with errorbar denotes the extracted critical filling at respective temperature. Dashed line denotes the BKT line $2T/\pi$.
}
\end{figure*}


We extract the superconducting $T_c^{\rm{SC}}$ for $|\JH|/\JA=0.26$ at a fixed $\nu$ through BKT criterion for $h_{\tn{SOC}}=0$, as shown in Fig.~\ref{Fig::ja_hsoc0}. The filling fractions are fixed through linear interpolation. The superconducting critical filling $\nu_c$ at a fixed inverse temperature $\beta$ are extracted through scaling analysis of $\xi_{\rm{SC}}/L$ for $h_{\tn{SOC}}=0.01$, as shown in Fig.~\ref{Fig::ja_hsoc001}.

\begin{figure*}[htb]
\includegraphics[width=180mm,scale=1]{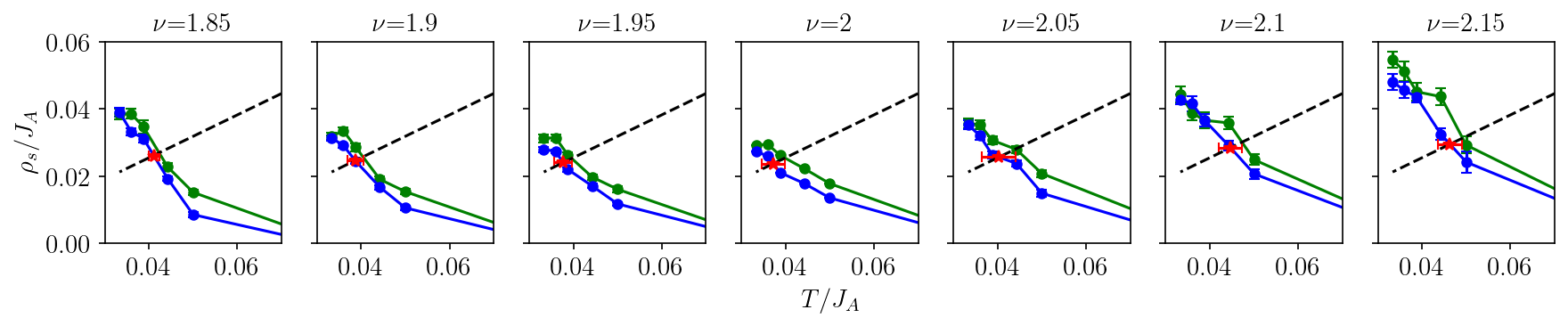}
\caption{\label{Fig::ja_hsoc0} 
Superfluid stiffness $\rho_s$ as a function of varying $T$ at a fixed filling $\nu$, calculated through Eq.~\ref{eq::stiffness} for $\mathcal{F}=0.01$, $|\JH|/\JA=0.26$ and $h_{\tn{SOC}}=0$. Filling fractions are fixed through linear interpolation. Color schemes are the same as that in Fig.~\ref{Fig::skyr_fig2}a. Red star with errorbar denotes the extracted $T_c^{\rm{SC}}$ at respective filling. Dashed line denotes the BKT line $2T/\pi$.
}
\end{figure*}

\begin{figure*}[htb]
\includegraphics[width=180mm,scale=1]{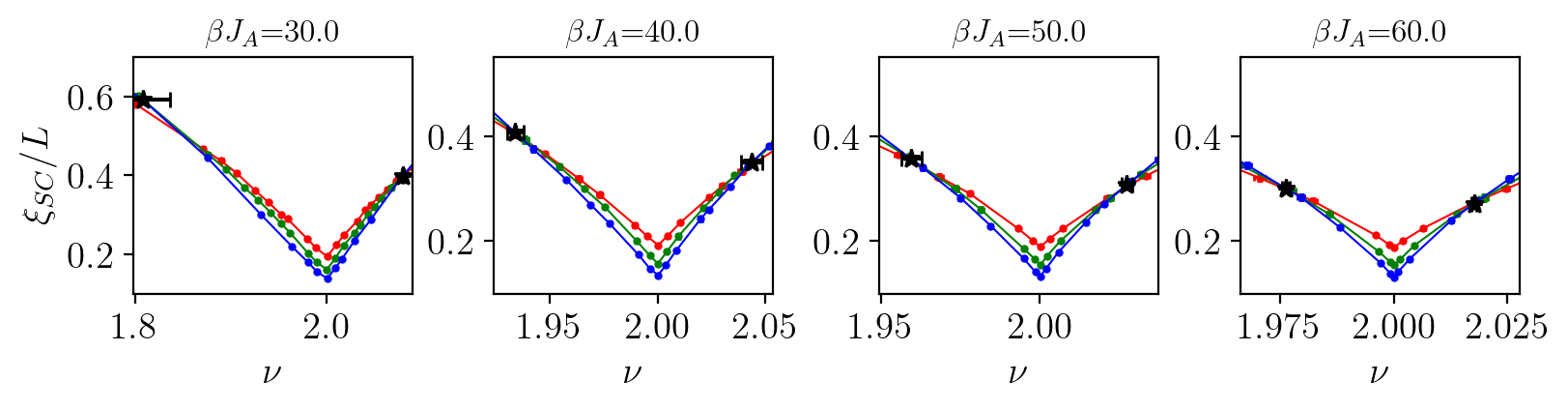}
\caption{\label{Fig::ja_hsoc001} 
Superconducting correlation length $\xi_{\rm{SC}}$ as a function of $\nu$ for $\mathcal{F}=0.01$, $|\JH|/\JA=0.26$ and $h_{\tn{SOC}}=0.01$. Black star with errorbar denotes the critical filling $\nu_c$ at the respective temperature. Color schemes are the same as that in Fig.~\ref{Fig::skyr_fig2}a.
}
\end{figure*}

\section{Low Temperature behavior of superfluid stiffness}\label{App::superlow}

In this section, we investigate and contrast the low-temperature behavior of the superfluid stiffness in the lightly-doped ``skyrmion-mediated" superconductor and the relatively heavily-doped superconductor (where the skyrmions evolve into electrons). We focus on $|\JH|/\JA=0.5$ and $\mathcal{F}=0.01$ with $h_{\rm{SOC}}=0$, where we have previously established the existence of skyrmions. We pick $\nu=2.04$ to probe the skyrmion-mediated superconductor and $\nu=2.16$ for analyzing the superconductor originating from condensing electronic Cooper pairs.

In Fig.~\ref{Fig::superlow}(a) and Fig.~\ref{Fig::superlow}(b), we show the superfluid stiffness $\rho_s(T)$ at $\nu=2.04$ on a semi-log scale and log-log scale, respectively; we show $\rho_s(T)$ at $\nu=2.16$ on a semi-log scale in Fig.~\ref{Fig::superlow}(c). At $\nu=2.16$, $\rho_s(T)$ saturates exponentially as a function of temperature, which aligns with the expectation for an ordinary fully-gapped superconductor. In contrast, the low-temperature behavior of $\rho_s(T)$ deviates from a clear exponential saturation (Fig.~\ref{Fig::superlow}a), which is likely due to the existence of gapless fluctuations tied to the CAF order. For the log-log plot of $\rho_s(T)-\rho_s(T=0)$ in Fig.~\ref{Fig::superlow}(b), $\rho_s(T=0)$ is obtained from an extrapolation using a polynomial function. The behavior of $\rho_s(T)$ at $\nu=2.04$ shows a tendency towards power-law saturation at low temperature. However, to determine the exact scaling, $\rho_s(T=0)$ should be calculated independently from zero-termperature projective Monte-Carlo simulation, which we leave for future study.

\begin{figure*}[htb]
\includegraphics[width=180mm,scale=1]{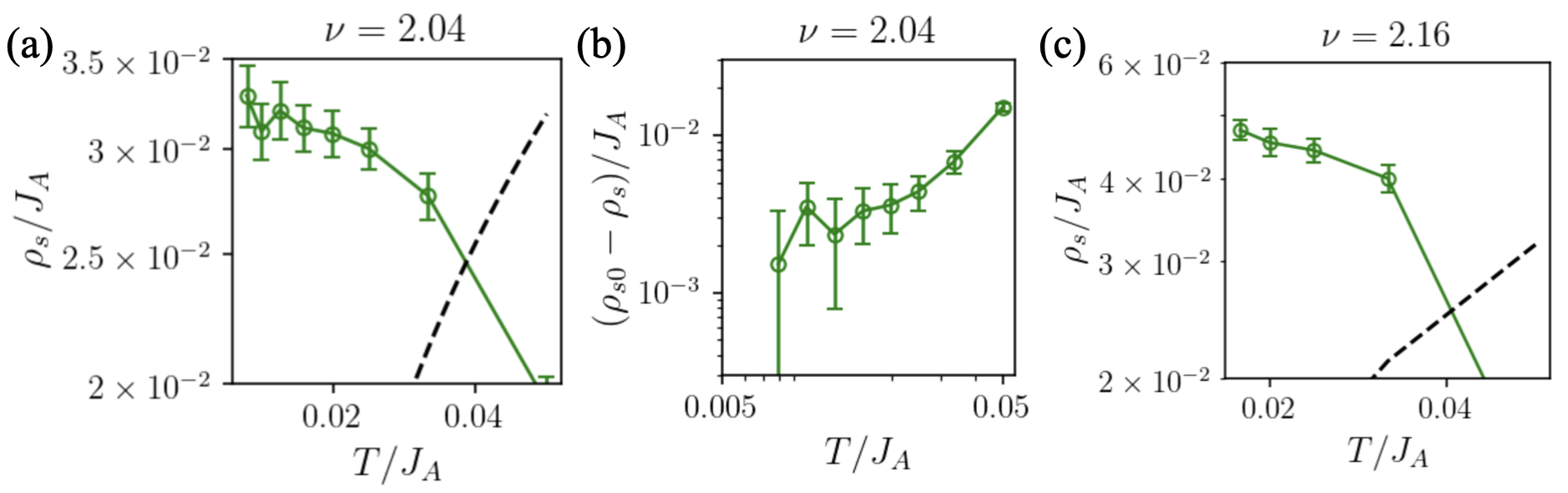}
\caption{\label{Fig::superlow} 
Saturation of superfluid stiffness at low temperature (a) at $\nu=2.04$ on a semi-log scale; (b) at $\nu=2.04$ on a log-log scale; (c) at $\nu=2.16$ on a semi-log scale.  Dashed lines are the BKT lines $2T/\pi$.
}
\end{figure*}

\end{widetext}

\end{document}